\documentclass[journal]{IEEEtran}

\usepackage{bm}
\usepackage{cite}
\usepackage{graphicx}
\usepackage{amsmath}
\usepackage{amssymb}
\usepackage{amsthm}
\usepackage{algorithm}
\usepackage{algorithmic}
\usepackage{array}
\usepackage{color}
\usepackage{multirow}
\usepackage{url}
\usepackage{soul}

\begin{document}
\title{Towards Intelligent Vehicular Networks: A Machine Learning Framework}

\author{
Le~Liang,~\IEEEmembership{Student Member,~IEEE}, Hao Ye,~\IEEEmembership{Student Member,~IEEE}, and\\
Geoffrey Ye Li,~\IEEEmembership{Fellow,~IEEE}
\thanks{
This work was supported in part by a research gift from Intel Corporation and in part by the National Science Foundation under Grants 1443894 and 1731017.
}
\thanks{
L. Liang, H. Ye and G. Y. Li are with the School of Electrical and Computer Engineering, Georgia Institute of Technology, Atlanta, GA 30332 USA (e-mail: \{lliang,yehao\}@gatech.edu; liye@ece.gatech.edu).
}
}

\maketitle

\begin{abstract}
As wireless networks evolve towards high mobility and providing better support for connected vehicles, a number of new challenges arise due to the resulting high dynamics in vehicular environments and thus motive rethinking of traditional wireless design methodologies.
Future intelligent vehicles, which are at the heart of high mobility networks, are increasingly equipped with multiple advanced onboard sensors and keep generating large volumes of data.
Machine learning, as an effective approach to artificial intelligence, can provide a rich set of tools to exploit such data for the benefit of the networks.
In this article, we first identify the distinctive characteristics of high mobility vehicular networks and motivate the use of machine learning to address the resulting challenges. After a brief introduction of the major concepts of machine learning, we discuss its applications to learn the dynamics of vehicular networks and make informed decisions to optimize network performance.
In particular, we discuss in greater detail the application of reinforcement learning in managing network resources as an alternative to the prevalent optimization approach.
Finally, some open issues worth further investigation are highlighted.
\end{abstract}

\begin{IEEEkeywords}
Machine learning, vehicular networks, high mobility, Internet of intelligent vehicles.
\end{IEEEkeywords}

\section{Introduction}
\begin{figure}[!t]
\centering
\includegraphics[width=\linewidth]{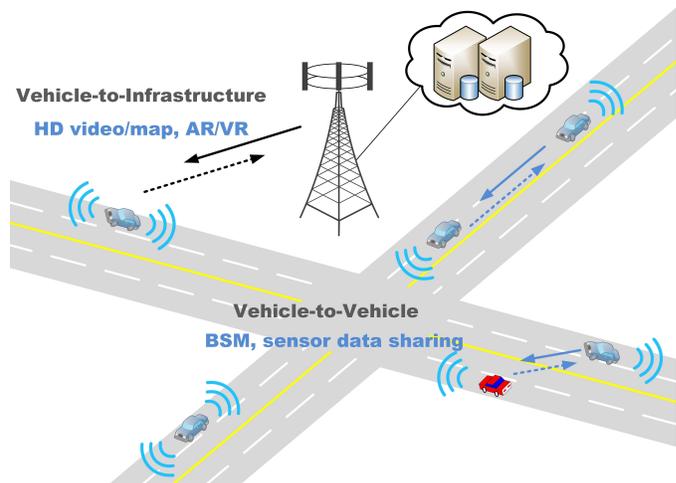}
\caption{An illustrative structure of vehicular networks.
\textcolor{black}{{HD: high-definition; AR: augmented reality; VR: virtual reality; BSM: basic safety message}}.}\label{fig:sysFig}
\end{figure}

Wireless networks that can support high mobility broadband access have received more and more attention from both industry and academia in recent years \cite{Liang2017vehicular,Peng2017vehicular,Araniti2013lte,Cheng2017Benz}.
In particular, the concept of connected vehicles or vehicular networks, as shown in Fig.~\ref{fig:sysFig}, has gained substantial momentum to bring a new level of connectivity to vehicles and, along with novel onboard computing and sensing technologies, serve as a key enabler of intelligent transportation systems (ITS) and smart cities \cite{Cheng2015D2D,Zhang2015novel}.
This new generation of networks will ultimately have a profound impact on the society, making everyday traveling safer, greener, and more efficient and comfortable.
Along with recent advances in a wide range of artificial intelligence (AI) technologies, it is helping pave the road to autonomous driving in the advent of the fifth generation cellular systems (5G).

Over the years, several communication standards for vehicular ad hoc networks (VANETs) have been developed across the globe, including dedicated short range communications (DSRC) in the United States \cite{Kenney2011dedicated} and the ITS-G5 in Europe \cite{ITSG5}, both based on the IEEE 802.11p technology \cite{ieee2010ieee}.
However, these technologies have been shown in recent studies \cite{Araniti2013lte,Hassan2011performance} to suffer from several issues, such as unbounded channel access delay, lack of quality of service (QoS) guarantees, and short-lived vehicle-to-infrastructure (V2I) connection.
To address the limitations of IEEE 802.11p based technologies and leverage the high penetration rate of cellular networks, the 3rd Generation Partnership Project (3GPP) has started to investigate supporting vehicle-to-everything (V2X) services in the long term evolution (LTE) network and the future 5G cellular system \cite{3GPPr14v2x,3GPPr15v2x}.
Some recent works along this line of effort can be found in \cite{Liang2017resource,Liang2017spectrum,Sun2016radio,Sun2016cluster,Botsov2014location,Cheng2015D2D}, which study efficient radio resource allocation for vehicular networks that employ the device-to-device (D2D) communications technology to support vehicle-to-vehicle (V2V) transmission in cellular systems.
\textcolor{black}{{ In addition, graph theoretic tools have been studied extensively for resource allocation design in vehicular networks \mbox{\cite{Zhang2013interference,Liang2018graph}} in recent years.}}
Major challenges in designing wireless networks to provide reliable and efficient support for high mobility environments result from the stringent and heterogeneous QoS requirements of vehicular applications as well as the strong dynamics that are inherent of the vehicular environment.

In the meantime, future intelligent vehicles are increasingly equipped with a wide variety of sensors, such as engine control units, radar, light detection and ranging (LIDAR), and cameras, to help the vehicle perceive the surrounding environment as well as monitoring its own operation status in real time.
Together with high performance computing and storage devices onboard, these sensing technologies are transforming vehicles from a simple transportation facility to a powerful computing and networking hub with intelligent processing capabilities. They keep collecting, generating, storing, and communicating large volumes of data, subject to further processing and commonly referred to as mobile big data \cite{cheng2017exploiting,Cheng2017mobile,Xu2018internet}.
Such data provide rich context information regarding the vehicle kinetics (such as speed, acceleration, and direction), road conditions, traffic flow, wireless environments, etc., that can be exploited to improve network performance through adaptive data-driven decision making.
However, traditional communications strategies are not designed to handle and exploit such information.

As a prevailing approach to AI, machine learning, in particular deep learning, has drawn considerable attention in recent years due to its astonishing progress in such areas as image classification \cite{krizhevsky2012imagenet}, video game playing \cite{mnih2015human}, and Go \cite{silver2016mastering}. It helps build intelligent systems to operate in complicated environments and has found many successful applications in computer vision, natural language processing, and robotics \cite{DLIntro,MLIntro}. Machine learning develops efficient methods to model and analyze large volumes of data by finding patterns and underlying structures and represents an effective data-driven approach to problems encountered in various scientific fields where heterogeneous types of data are available for exploitation. As a result, machine learning provides a rich set of tools that can be leveraged to exploit the data generated and stored in vehicular networks \cite{Jiang2017machine,Ye2018machine} and help the network make more informed and data-driven decisions. However, how to adapt and exploit such tools to account for the distinctive characteristics of high mobility vehicular networks and serve the purpose of reliable vehicular communications remains challenging and represents a promising research direction.

\textcolor{black}{{In this paper, we identify and discuss major challenges in supporting vehicular networks with high mobility, such as fast-varying wireless channels, volatile network topologies, ever-changing vehicle densities, and heterogeneous QoS requirements for diverse vehicular links.
To address these challenges, we deviate from the traditional network design methodology and motivate the use of various machine learning tools, including a range of supervised, unsupervised, deep, and reinforcement learning methods, to exploit the rich sources of data in vehicular networks for the benefits of communications performance enhancement.
In particular, we discuss in greater detail recent advances of leveraging machine learning to acquire and track the dynamics of vehicular environments, automatically make decisions regarding vehicular network traffic control, transmission scheduling and routing, and network security, and perform intelligent network resource management based on reinforcement learning techniques.
Since research in this area is still in its infancy, a wide spectrum of interesting research problems are yet to be defined and fully explored. We list a few of them in this paper and hope to bring more attention to this emerging field.}}

The rest of this paper is organized as follows.
In Section II, we introduce the unique characteristics and challenges of high mobility vehicular networks and motivate the use of machine learning to address the challenges.
In Section III, we discuss the basic concepts and major categories of machine learning, and then investigate how to apply machine learning to learn the dynamics of high mobility networks in Section IV.
In Section V, we present some preliminary examples of applying machine learning for data-driven decision making and wireless resource management problems in vehicular networks.
In Section VI, we recognize and highlight several open issues that warrant further research and concluding remarks are finally made in Section VII.

\section{Challenges of High Mobility Vehicular Networks}
High mobility vehicular networks exhibit distinctive characteristics, which have posed significant challenges to wireless network design. In this section, we identify such challenges and then discuss the potential of leveraging machine learning to address them.

\subsection{Strong Dynamics}\label{sec_strong}
High mobility of vehicles leads to strong dynamics and affects system design in multiple aspects of the communications network.
Special channel propagation characteristics are among the most fundamental differentiating factors of high mobility networks compared with low mobility counterparts.
For example, vehicular channels exhibit rapid temporal variation and also suffer from inherent non-stationarity of channel statistics due to their unique physical environment dynamics \cite{Viriyasitavat2015vehicular,Bernado2014delay}.
Such rapid variations induce short channel coherence time and bring significant challenges in acquiring accurate channel estimates at the receiver in real time. This is further hindered by the non-stationarity of channel statistics, which are usually leveraged to improve estimation accuracy \cite{Li1998robust,Li2000pilot,ding2015sparse}.
Meanwhile, due to the high Doppler spread caused by vehicle mobility, the multicarrier modulation scheme is more susceptible to intercarrier interference (ICI) in vehicular networks \cite{Russell1995interchannel,Li2001bounds} and hence brings difficulty to signal detection.
Constant mobility of vehicles also causes frequent changes of the communications network topology, affecting channel allocation and routing protocol designs. For example, in cluster-based vehicular networks \cite{Abboud2014stochastic}, moving vehicles may join and leave the cluster frequently, making it hard to maintain long-lasting connections within the formed cluster and thus warranting further analysis on cluster stability.
Another source of dynamics in high mobility networks comes from the changing vehicle density, which varies dramatically depending on the locations (remote suburban or dense urban areas) and time (peak or off hours of the day). Flexible and robust resource management schemes that make efficient use of available resources while adapting to the vehicle density variation are thus needed.

Traditionally developed rigorous mathematical theories and methods for wireless networks are mostly based on static or low-mobility environment assumptions and usually not designed to treat the varying environment conditions in an effective way. Therefore, it is important to explore new methodologies that can interact with fast changing environments and obtain optimal policies for high mobility vehicular networks in terms of both physical layer problems, such as channel estimation and signal detection and decoding, and upper layer designs, such as resource allocation, link scheduling and routing.

\subsection{Heterogeneous and Stringent QoS Requirements}

In high mobility vehicular networks, there exist different types of connections, which we broadly categorize into V2I and V2V links.
The V2I links enable vehicles to communicate with the base station to support various traffic efficiency and information and entertainment (infotainment) services. They generally require frequent access to the Internet or remote servers for media streaming, \textcolor{black}{{high-definition (HD) map downloading}}, and social networking, which involve considerable amount of data transfer and thus are more bandwidth intensive \cite{Araniti2013lte}.
On the other hand, the V2V links are mainly considered for sharing safety-critical information, \textcolor{black}{{as basic safety messages (BSM) in DSRC}} \cite{Kenney2011dedicated}, among vehicles in close proximity in either a periodic or event triggered manner.
Such safety related messages are strictly delay sensitive and require very high reliability. For example, the European METIS project requires the end-to-end latency to be less than 5 milliseconds and the transmission reliability to be higher than 99.999\% for a safety packet of 1600 bytes \cite{METIS2013}.
Moreover, high bandwidth sensor data sharing among vehicles is currently being considered in 3GPP for V2X enhancement in future 5G cellular networks for advanced safety applications \cite{3GPPr15v2x}, whose quality degrades gracefully with increase in packet delay and loss.
As a result, stringent QoS requirements of low latency and high reliability are in turn imposed on the V2V links.
Traditional wireless design approaches are hard to simultaneously meet such diverse and stringent QoS requirements of vehicular applications, which is further challenged by the strong dynamics in high mobility vehicular networks as discussed in Section~\ref{sec_strong}.

\subsection{The Potential of Machine Learning}
Machine learning emphasizes the ability to learn and adapt to the environment with changes and uncertainties.
Different from the traditional schemes that rely on explicit system parameters, such as the received signal power or signal-to-interference-plus-noise ratio (SINR), for decision making in vehicular networks, machine learning can exploit multiple sources of data generated and stored in the network (e.g., power profiles, network topologies, vehicle behavior patterns, the vehicle locations/kinetics, etc.) to learn the dynamics in the environment and then extract appropriate features to use for the benefit of many tasks for communications purposes, such as signal detection,  resource management, and routing.
However, it is a non-trivial task to extract context or semantic information from a huge amount of accessible data, which might have been contaminated by noise, multi-modality, or redundancy, and thus information extraction and distillation need to be performed.

In particular, reinforcement learning \cite{sutton1998introduction}, one of the machine learning tools, can interact with the dynamic environment and develop satisfactory policies to meet diverse QoS requirements of vehicular networks while adapting to the varying wireless environment.
For example, in resource allocation problems, the optimal policies are first learned and then the vehicle agents accordingly take actions to adjust powers and allocate channels adaptive to the changing environments characterized by, e.g, link conditions, locally perceived interference, and vehicle kinetics while traditional static mathematical models are not good at capturing and tracking such dynamic changes.

\section{Machine Learning}
Machine learning allows computers to find hidden insights through iteratively learning from data, without being explicitly programmed. It has revolutionized the world of computer science by allowing learning with large datasets, which enables machines to change, re-structure and optimize algorithms by themselves.
Existing machine learning methods can be divided into three categories, namely, supervised learning, unsupervised learning, and reinforcement learning.
Other learning schemes, such as semi-supervised learning, online learning, and transfer learning, can be viewed as variants of these three basic types. In general, machine learning involves two stages, i.e., training and testing. In the training stage, a model is learned based on the training data while in the testing stage, the trained model is applied to produce the prediction.
In this section, we briefly introduce the basics of machine learning in the hope that the readers can appreciate their potential in solving traditionally challenging problems.


\subsection{Supervised Learning}

The majority of practical machine learning algorithms use supervised learning with a labeled dataset, where each training sample comes with a label. The ultimate goal of supervised learning is to find the mapping from the input feature space to the label so that  reliable prediction can be made when new input data is given.
Supervised learning problems can be further categorized into classification and regression, where the difference between the two tasks is that the labels are categorical for classification and numerical for regression.

Classification algorithms learn to predict a category output for each incoming sample based on the training data. Some classic algorithms in this category include Bayesian classifiers \cite{box2011bayesian}, $k$-nearest neighbors (KNN) \cite{beyer1999nearest}, decision trees \cite{safavian1991survey}, support vector machine (SVM) \cite{cortes1995support}, and neural networks \cite{lecun2015deep}.
Instead of discrete outputs, regression algorithms predict a continuous value corresponding to each sample, such as estimating the house price given its associated feature inputs. Classic regression algorithms include logistic regression \cite{walker1967estimation}, support vector regression (SVR) \cite{basak2007support}, and Gaussian process for regression \cite{box2011bayesian}.

\subsection{Unsupervised Learning}

The label data serves as the teacher in supervised learning so that there is a clear measure of success that can be used to judge the goodness of the learned model in various situations.
Nevertheless, a large amount of labeled data is often hard to obtain in practice. As a consequence, learning with unlabeled data, known as unsupervised learning, has been developed to find an efficient representation of the data samples without any labeling information. For instance, samples might be explained by hidden structures or hidden variables, which can be represented and learned by Bayesian learning methods.

A representative case of unsupervised learning is clustering, namely, to group samples in a way that samples in the same cluster have more similarities than the samples in different clusters. The features used for clustering could be either the absolute description of each sample or the relative similarities between samples.
Classic clustering algorithms include $k$-means \cite{kanungo2002efficient}, hierarchical clustering \cite{gan2007data}, spectrum clustering \cite{ng2002spectral}, and Dirichlet process \cite{teh2011dirichlet}.

Besides clustering,  dimension reduction is another important case of unsupervised learning, where samples from a high dimensional space are projected into a lower one without losing too much information.
In many scenarios, the raw data come with high dimension, which is not desirable because of several reasons. One reason is the so-called \emph{curse of dimensionality} \cite{friedman1997bias}, which describes the problematic phenomenon encountered when the dimension becomes huge. For instance, in optimization, clustering, and classification, the model complexity and the number of required training samples grow dramatically with the feature dimension.
Another reason is that the inputs of each dimension are usually correlated and some dimensions may be corrupted with noise and interference, which would degrade the learning performance if not handled properly.
Some classic dimension reduction algorithms include linear projection methods, such as principal component analysis (PCA) \cite{jolliffe1986principal}, and nonlinear projection methods, such as manifold learning, local linear embedding (LLE) \cite{roweis2000nonlinear}, and isometric feature mapping (ISOMAP) \cite{tenenbaum2000global}.

\subsection{Reinforcement Learning}
In reinforcement learning problems, an agent learns the optimal behaviors through interacting with the environment in a trial-and-error manner aiming to maximize rewards from the environment. The environment is modeled as a Markov decision process (MDP), which introduces actions and rewards to a Markov process.
Both the state transition probability, $p(s',r|s, a)$, and the reward, $r$, are determined only by the current state, $s$, and the selected action, $a$.
The goal of reinforcement learning is to find a policy that takes the action to maximize the future discounted reward, defined as
\begin{align}
G_t = R_{t+1} + \gamma R_{t+2} + \gamma^2 R_{t+3} ... = R_{t+1} + \gamma G_{t+1},
\end{align}
where $\gamma$ is the discount factor and $R_t$ is the reward at each time step $t$ \cite{sutton1998introduction}.

Learning the $Q$ function is a classic approach to solve the reinforcement learning problem, where the $Q(s, a)$ function estimates the expectation of the sum reward when taking an action $a$, in a given state $s$. The optimal $Q$ function is the maximum expected sum reward achievable by following any policy of choosing actions and constrained by the Bellman equation,
\begin{align}
Q^*(s,a) =\sum_{s',r}p(s',r|s,a)[r+\gamma\max_{a' \in \mathcal{A}}Q^*(s',a')],
\end{align}
where $A$ is the action set.
In general, at the fixed point of Bellman equation, the optimal $Q$ function can be found by performing iterative updates, after which the the optimal policy can be determined by taking the action that maximizes the $Q$ function.
A rich set of algorithms, such as Sarsa \cite{sutton1998introduction} and Q-learning \cite{watkins1989learning}, have been historically designed to serve such purposes.
Reinforcement learning can be applied in vehicular networks to handle the temporal variation of wireless environments, which will be discussed in Section~\ref{sec:resource} in detail.

\subsection{Deep Learning} \label{sec:dlIntro}

\begin{figure}[!t]
\centering
\includegraphics[width=\linewidth]{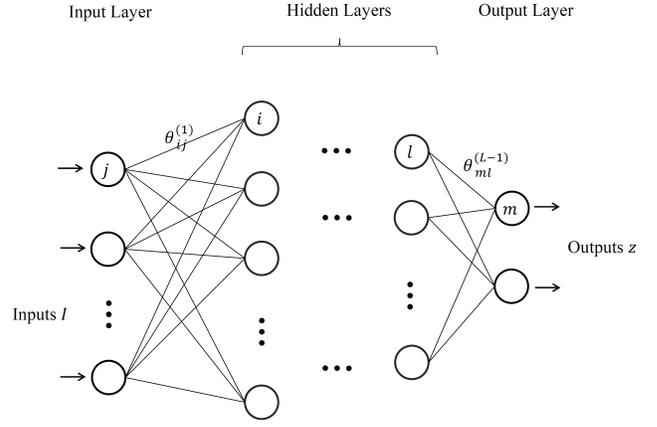}
\caption{An illustration of deep neural networks with multiple layers.}\label{fig:dl}
\end{figure}

\begin{table*}[t]
\centering
\textcolor{black}{
\caption{{{\normalsize Summary of dynamics learning in vehicular networks}}.}\label{tab:dynamics}
\normalsize
\begin{tabular}{|l|l|l|}
\hline
\textbf{Tasks}      & \textbf{Learning Methods}               & \textbf{References}       \\ \hline
\multicolumn{1}{|c|}{\multirow{2}{*}{Learning-based channel estimation}} & Bayesian learning              & \cite{wen2015channel}         \\ \cline{2-3}
\multicolumn{1}{|c|}{}                                                     & Deep learning                  & \cite{Ye2018power}         \\ \hline
\multirow{2}{*}{Traffic flow prediction}                                   & Deep learning                  & \cite{lv2015traffic}        \\ \cline{2-3}
                                                                           & Probabilistic graphical models & \cite{ide2015lte}       \\ \hline
Vehicle trajectory prediction                                              & Gaussian mixture models        & \cite{Wiest2012probabilistic}         \\ \hline
\end{tabular}
}
\end{table*}

Deep learning aims to learn data representations, which can be built in supervised, unsupervised, and reinforcement learning and has made significant advances in various machine learning tasks.
As a deeper version of neural networks, which consist multiple layers of neurons, the structure of deep learning is shown in Fig.~\ref{fig:dl}. The input layer is at the left, where each node in the figure represents a dimension of the input data, while the output layer is at the right, corresponding to the outputs. The layers in the middle are called hidden layers.
Each neuron in the network performs a non-linear transform on a weighted sum of a subset of neurons in its preceding layer.
The nonlinear function may be the Sigmoid function, or the Relu function, defined as
$f_{S}(a) = \frac{1}{1+e^{-a}}$, and $f_{R}(a) = \max(0,a)$, respectively.
Hence, the output of the network $\mathbf{z}$ is a cascade of nonlinear transform of the input data $\mathbf{I}$, mathematically expressed as
\begin{equation}
\mathbf{z} = f(\mathbf{I}, \boldsymbol{\theta}) = f^{(L-1)}(f^{(L-2)}(\cdot \cdot \cdot f^{(1)}(\mathbf{I}))),
\end{equation}
where $L$ is the layer index and $\boldsymbol{\theta}$ denotes the weights of the neural network.
Typically, the neural network's representation ability grows as the hidden layers become deeper. However, numerous barriers occur when training deeper networks, such as much more training data is needed and gradients of networks may easily explode or vanish \cite{DLIntro}.
By virtue of development in  faster computation resources, new training methods (new activation functions \cite{glorot2011deep}, pre-training \cite{erhan2010does}), and new structures (batch norm \cite{ioffe2015batch}, residual networks \cite{he2016deep}), training a much deeper neural network becomes viable.
Recently, deep learning has been widely used in computer vision \cite{krizhevsky2012imagenet}, speech recognition \cite{weng2014recurrent}, natural language processing \cite{cho2014learning}, etc., and has greatly improved state-of-the-art performance in each area.
In addition, different structures can be added to the deep neural networks for different applications. For example, convolutional networks share weights among spatial dimensions while recurrent neural networks (RNN) and long short term memory (LSTM) networks share weights among the temporal dimensions \cite{DLIntro}.

\section{Learning Dynamics}\label{sec_dynamics}

High mobility networks exhibit strong dynamics in many facets, e.g., wireless channels, network topologies, traffic dynamics, etc., that heavily influence the network performance.
In this section, we discuss how to exploit machine learning to efficiently learn and robustly predict such dynamics based on data from a variety of sources.
Table~\ref{tab:dynamics} summarizes these tasks along with the leveraged machine learning methods.

\subsection{Learning-Enabled Channel Estimation}
Accurate and efficient channel estimation is a critical component in modern wireless communications systems. It has strong impacts on receiver design (e.g., channel equalization, demodulation, decoding, etc.) as well as radio resource allocation at the transmitter for interference mitigation and performance optimization. Channel estimation is more of an issue in vehicular networks with high Doppler shifts and short channel coherence periods.

Statistical information of wireless channels, such as time and frequency domain correlation, mainly depends on vehicle locations/speeds, multipath delay spread, and the surrounding wireless environment.
In cellular based vehicular networks, the base station can easily access accurate location information (speed can also be inferred) of all vehicles traveling under its coverage from various global navigation satellite systems (GNSS) on vehicles.
It can maintain a dynamic database to store the historical estimates of communications channels for all vehicular links along with relevant context information, such as locations of the transmitters and/or receivers and traffic patterns.
Various machine learning tools, such as Bayesian learning and deep learning, can then be leveraged to exploit such historical data to predict the channel statistics and enhance instantaneous channel estimation for current vehicular links.

\textcolor{black}{{Compared with traditional channel estimation schemes relying on precise mathematical models, the learning-based method provides yet another data-driven approach that can easily incorporate various sources of relevant context information to enhance estimation accuracy. It can potentially deal with a number of non-ideal effects that are difficult to handle under the traditional estimation framework, such as nonlinearity of power amplifiers, phase noise, and time/frequency offsets. The channel estimator can be trained offline across different channel models for varying propagation environments and calibrated using real-world data collected from field measurements. During online deployment, the estimator module produces channel estimates on the fly with low computational complexity given necessary inputs, which includes received pilot data and other relevant context information.}}

For example, a Bayesian learning approach has been adopted to estimate the sparse massive multiple-input multiple-output (MIMO) channel in \cite{wen2015channel}, where the channel is modeled using Gaussian mixture distribution and an efficient estimator has been derived based on approximate message passing (AMP) and expectation-maximization (EM) algorithms.
Deep learning has been exploited in \cite{Ye2018power} to implicitly estimate wireless channels in orthogonal frequency division multiplexing (OFDM) systems and shown to be robust to nonlinear distortions and other impairments, such as pilots reduction and cyclic prefix (CP) removal.
In addition, the temporal relationship in data is traditionally characterized by Bayesian models, such as the HMMs, which can be used to track time-varying vehicular channels. It is interesting to see if recently developed sophisticated models powered by deep neural networks, such as RNN and LSTM, can improve channel estimation accuracy by exploiting the long-range dependency.

\subsection{Traffic Flow Prediction}
Traffic flow prediction aims to infer traffic information from historical and real-time traffic data collected by various onboard and roadway sensors.
It can be used in a variety of ITS applications, such as traffic congestion alleviation, fuel efficiency improvement, and carbon emission reduction.
Given the rich amount of traffic data, machine learning can be leveraged to enhance the flow prediction performance and achieve unparalleled accuracy.
In \cite{lv2015traffic}, a deep learning based method has been proposed to predict traffic flow, where a stacked autoencoder is exploited to learn generic features for traffic flow and trained in a greedy layerwise fashion. It implicitly takes into consideration the spatial and temporal correlations in the modeling and achieves superior performance.
A probabilistic graphical model, namely the Poisson dependency network (PDN), has been learned in \cite{ide2015lte} to describe empirical vehicular traffic dataset and then used for traffic flow prediction. The strong correlations between cellular connectivity and vehicular traffic flow have been further leveraged to enhance prediction for both of them by means of Poisson regression trees.

\subsection{Vehicle Trajectory Prediction}
Vehicle trajectory prediction is of significant interest for advanced driver assistance systems (ADAS) in many tasks, such as collision avoidance and road hazard warning.
It also plays an important role in networking protocol designs, such as handoff control, link scheduling, and routing, since network topology variations can be inferred from the predicted vehicle trajectories and exploited for communications performance enhancement.
Probabilistic trajectory prediction based on Gaussian mixture models (GMM) and variational GMM has been studied in \cite{Wiest2012probabilistic} to predict the vehicle's trajectory using previously observed motion patterns.
A motion model is learned based on previously observed trajectories, which is then used to build a functional mapping from the observed historical trajectories to the most likely future trajectory.
The latent factors that affect the trajectories, such as drivers' intention, traffic patterns, and road structures, may also be implicitly learned from the historical data using deep neural networks.
More sophisticated models, such as RNN and LSTM, can potentially lead to better results for modeling the dynamics of vehicle trajectories and are worth further investigation.

\begin{table*}[t]
\centering
\caption{{{\normalsize  Summary of learning based decision making in vehicular networks}}.}\label{tab:decision}
\normalsize
\begin{tabular}{|l|l|l|}
\hline
\textbf{Tasks}      & \textbf{Learning Methods}               & \textbf{References}       \\ \hline
\multirow{3}{*}{Location prediction based scheduling and routing}          & Hidden Markov models            & \cite{Yao2017v2x}         \\ \cline{2-3}
                                                                           & Variable-order Markov models   & \cite{Xue2012novel}         \\ \cline{2-3}
                                                                           & Recursive least squares        & \cite{zeng2017channel}         \\ \hline
Network congestion control                                                 & $k$-means clustering             & \cite{taherkhani2016centralized}         \\ \hline
Load balancing and vertical handoff control                                        & Reinforcement learning         & \cite{li2017user,xu2014fuzzy}     \\ \hline
\multirow{2}{*}{Network security}                                          & Deep learning                  & \cite{kang2016novel}         \\ \cline{2-3}
                                                                           & Long short term memory         & \cite{taylor2016anomaly}         \\ \hline
Virtual resource allocation                                                & (Deep) reinforcement learning  & \cite{zheng2016delay, salahuddin2016reinforcement, He2017integrated} \\ \hline
Energy-efficient resource management                                       & (Deep) reinforcement learning  & \cite{atallah2017reinforcement,atallah2017deep}     \\ \hline
Distributed resource management                                            & Deep reinforcement learning   &
\cite{Ye2018deep} \\ \hline
\end{tabular}
\end{table*}

\section{Learning Based Decision Making in Vehicular Networks}
The rich sources of data generated and stored in vehicular networks motivate a data-driven approach for decision making that is adaptive to network dynamics and robust to various impairments.
Machine learning represents an effective tool to serve such purposes with proven good performance in a wide variety of applications, as demonstrated by some preliminary examples discussed in this section and summarized in Table~\ref{tab:decision}.

\subsection{Location Prediction Based Scheduling and Routing}
We have shown in Section~\ref{sec_dynamics} that machine learning can be leveraged to learn the dynamics in high mobility vehicular networks, including vehicle trajectory prediction. In fact, the predicted dynamics can be further used towards networking protocol designs for system performance improvement.
For example, the hidden Markov model (HMM) has been applied in \cite{Yao2017v2x} to predict vehicles' future locations based on past mobility traces and movement patterns in a hybrid VANET with both V2I and V2V links.
Based on the predicted vehicle trajectories, an effective routing scheme has been proposed to efficiently select relay nodes for message forwarding and enable seamless handoff between V2V and V2I communications.
A variable-order Markov model has been adopted in \cite{Xue2012novel} to extract vehicular mobility patterns from real trace data in an urban vehicular network environment, which is used to predict the possible trajectories of moving vehicles and develop efficient prediction-based soft routing protocols.
In \cite{zeng2017channel}, a recursive least squares algorithm has been used for large-scale channel prediction based on location information of vehicles and facilitate the development of a novel scheduling strategy for cooperative data dissemination in VANETs.


\subsection{Network Congestion Control}
Data traffic congestion is an important issue in vehicular networks, especially when the network conditions are highly dense in, e.g., busy intersections and crowded urban environments. In such cases, a large number of vehicles are vying for the available communication channels simultaneously and hence cause severe data collisions with increased packet loss and delay.
To guarantee a reliable and timely delivery of various delay-sensitive safety-critical messages, such as BSMs, the vehicular networks need to have carefully designed congestion control strategies.
Traditionally, there are five major categories of congestion control methods, namely rate-based, power-based, carrier-sense multiple access/collision avoidance based, prioritizing and scheduling-based, and hybrid strategies \cite{taherkhani2016centralized}, which adjust communications parameters, such as transmission power, transmission rates, and contention window sizes, etc., to meet the congestion control purposes.

Different from the traditional approaches, an effective machine learning based data congestion control strategy utilizing $k$-means clustering has been developed in \cite{taherkhani2016centralized} for congestion prone intersections.
The proposed strategy relies on local road side units (RSUs) installed at each intersection for congestion detection, data processing, and congestion control to provide a centralized congestion management for all vehicles that are passing through or stop at the intersection.
After detection of congestion, each RSU collects all data transferred among vehicles in its coverage, removes their redundancy, exploits $k$-means algorithms to cluster the messages according to their features, such as sizes, validity, and types, and finally adjusts communications parameters for each cluster.

\subsection{Load Balancing and Vertical Handoff Control}
Due to periodicity of everyday traffic, potential patterns and regularities lie in the traffic flow and can be further exploited with learning based methods for load balancing and vertical control in vehicular networks.
An online reinforcement learning approach has been developed in \cite{li2017user} to address the user association problem with load-balancing in the dynamic environment. The initial association is achieved based on the current context information using reinforcement learning. After a period of learning, with the association information being collected at the base station, the new association results will be obtained directly and adaptively using historical association patterns.
Besides user association, the reinforcement learning based approach has also been applied in \cite{xu2014fuzzy} to the vertical handoff design for heterogeneous vehicular networks. The network connectivity can be determined by a fuzzy Q-learning approach with four types of information, namely, received signal strength value, vehicle speed, data quantity, and the number of users associated with the targeted network. With the learning based strategy, users can be connected to the best network without prior knowledge on handoff behavior.

\subsection{Network Security}\label{sec:networkSecurity}

As intelligent vehicles become more connected and bring huge benefits to the society, the improved connectivity can make vehicles more vulnerable to cyber-physical attacks.
As a result, security of information sharing in vehicles is crucial since any faulty sensor measurements may cause accidents and injuries.
In \cite{kang2016novel}, an intrusion detection system has been proposed for vehicular networks based on deep neural networks, where the unsupervised deep belief networks are used to initialize the parameters as a preprocessing stage. Then, the deep neural networks are trained by high-dimensional packet data to figure out the underlying statistical properties of normal and hacking packets and extract the corresponding features. In addition, LSTM is used in \cite{taylor2016anomaly} to detect attacks on connected vehicles. The LSTM based detector is able to recognize the synthesized anomalies with high accuracy by learning to predict the next word originating from each vehicle.

\subsection{Intelligent Wireless Resource Management}\label{sec:resource}
The current mainstream approach to wireless resource management is to formulate the design objective and constraints as an optimization problem and then solve for a solution with certain optimality claims.
However, in high mobility vehicular networks, such an approach is insufficient. The first challenge arises due to the strong dynamics in vehicular networks that lead to a brief valid period of the optimization results in addition to the incurred heavy signaling overhead.
The second issue comes with the difficulty to formulate a satisfactory objective to simultaneously consider the vastly different goals of the heterogeneous vehicular links, which is further complicated by the fact that some of the QoS formulations are mathematically difficult if not intractable.
Fortunately, reinforcement learning provides a promising solution to these challenges through interacting with the dynamic environment to maximize a numeric reward, which is discussed in detail in this part.

\subsubsection{Virtual Resource Allocation}


Employing recent advances in software-defined networking (SDN) and network function virtualization (NFV), the traditional vehicular network can be transformed into a virtualized network offering improved efficiency and greater flexibility in network management.
Future intelligent vehicles and RSUs will be equipped with advanced sensing, computing, storage, and communication facilities, which can be further integrated into the virtualized vehicular network to provide a pool of resources for a variety of ITS applications.
In such a complicated system, how to dynamically allocate the available resources to end users for QoS maximization with minimal overhead is a nontrivial task.
A delay-optimal virtualized radio resource management problem in software-defined vehicular networks has been considered in \cite{zheng2016delay}, which is formulated as an infinite-horizon partially observed MDP.
An online distributed learning algorithm has been proposed to address the problem based on an equivalent Bellman equation and stochastic approximation. The proposed scheme is divided into two stages, which adapt to large time scale factors, such as the traffic density, and small timescale factors, such as channel and queue states, respectively.
In \cite{salahuddin2016reinforcement}, the resource allocation problem in vehicular clouds has been modeled as an MDP and reinforcement learning is leveraged to solve the problem such that the resources are dynamically provisioned to maximize long-term benefits for the network and avoid myopic decision making.
Joint management of networking, caching, and computing resources in virtualized vehicular networks has been further considered in \cite{He2017integrated}, where a novel deep reinforcement learning approach has been proposed to deal with the highly complex joint resource optimization problem and shown to achieve good performance in terms of total revenues for the virtual network operators.

\subsubsection{Energy-Efficient Resource Management}
Energy consumption should be taken into consideration, especially when RSUs in vehicular networks lack permanent grid-power connection.
In \cite{atallah2017reinforcement}, an MDP problem is formulated and solved using reinforcement learning techniques to optimize the RSUs' downlink scheduling performance during a discharge period. The RSUs learn to select a vehicle to serve at the beginning of each time slot based on the collected information about traffic characteristics, infrastructure power budget, and the total length of a discharge period. The reward function is set as the performance metric for the total number of downloaded bits and the number of fulfilled vehicle requests per discharge period.
Q-learning is then employed to solve the problem and obtain the highest reward in the long run. The framework can be further extended and augmented by deep reinforcement learning as in \cite{atallah2017deep}, where a deep reinforcement learning based scheduling scheme has been proposed that can overcome the drawback of using discrete states and actions.
It first performs random scheduling policy and then gradually learns an adaptive dynamic policy to extend the battery life, minimize the reception latency, and achieve the QoS levels. Deep reinforcement learning augments the RSU with the ability to observe and analyze the environment and make decisions.

\subsubsection{Distributed Resource Management}
\begin{figure}[!t]
\centering
\includegraphics[width = \linewidth]{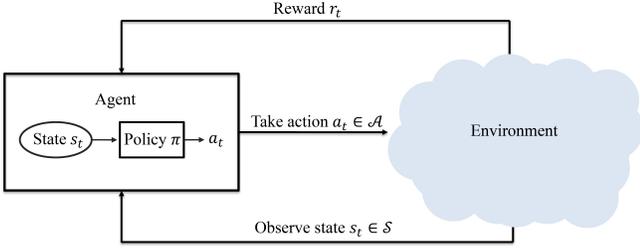}
\caption{An illustration of deep reinforcement learning for resource management in vehicular networks.} \label{fig:rl}
\end{figure}

Most of resource allocation algorithms for D2D-based vehicular networks are conducted in a centralized manner, where the central controller collects information and makes decisions for all the vehicles by solving optimization problems. However, In order to acquire the global network information, centralized control schemes will incur huge overhead, which grows dramatically with the size of vehicular networks. As shown in Fig.~\ref{fig:rl}, we have proposed a deep reinforcement learning based decentralized resource allocation mechanism for vehicular networks  \cite{Ye2018deep}, where the mapping from the partial observations of each vehicle agent to the optimal resource allocation solution can be approximated by deep neural networks. The merit of reinforcement learning based method is that it can address stringent latency requirements on V2V links, which is usually hard to deal with using existing optimization approaches.

The V2I link is assumed to have been allocated orthogonal resources beforehand and the main goal of the proposed distributed spectrum and power allocation is to satisfy the latency constraints for each V2V link and minimize interference to V2I links.
The structure of reinforcement learning for V2V communications is shown in Fig.~\ref{fig:rl}, where an agent, corresponding to a V2V link, interacts with the environment.
In this scenario, the environment is considered to be everything beyond the V2V link. Since the behavior of other V2V links is controlled by other agents in the decentralized settings, their actions are treated as part of the environment.

As shown in Fig.~\ref{fig:rl}, at time $t$, each an agent, i.e., each V2V link, observes a state, $s_t$, from the state space, $\mathcal{S}$, and accordingly takes an action, $a_t$, selected from the action space, $\mathcal{A}$,  which amounts to selecting the sub-band and transmission power based on the policy, $\pi$.
The decision policy, $\pi$, is determined by a Q-function, $Q(s_t, a_t, \theta)$, where $\theta$ is the parameter of the Q-function.
With actions taken, the environment transitions to a new state, $s_{t+1}$, and the agent receives a reward, $r_t$, which is determined by the capacity of the V2I link and V2V link as well as the corresponding latency.
The state observed by each V2V link consists of several components: the instantaneous channel information of the corresponding V2V link, $g_t$, the previous interference to the link, $I_{t-1}$, the channel information of the V2I link, $h_t$, the selection of sub-bands of neighbors in the previous time slot, $B_{t-1}$, the remaining load for the vehicles to transmit, $L_t$, and the remaining time to meet the latency constraints $U_t$. Hence the state can be expressed as $s_t = [g_t, I_{t-1}, h_t, B_{t-1}, L_t, U_t]$. The instantaneous channel information and the received interference relate to the quality of each sub-band. The distribution of neighbors' selection reveals the interference to other vehicles. In addition, the remaining amount of messages to transmit and the remaining time could be useful for selecting suitable power levels.
Q-learning is employed to obtain an optimal policy for resource allocation in V2V communications to maximize the long-term expected accumulated discounted rewards, $G_t$, where the $Q$ function is approximated by a deep neural network. The optimal policy with Q-values $Q^*$ can be found without any knowledge of the underlying system dynamics based on the following update equation,
\begin{equation}
\begin{aligned}
Q_{new}({s_t}, a_t)   = & Q_{old}(\mathbf{s_t}, a_t) + \alpha [r_{t+1} \nonumber\\
& +\gamma \max_{{s}\in \mathcal{S}} Q_{old}({s}, a_t) - Q_{old}({s_t}, a_t)].
\end{aligned}
\end{equation}
The training and testing samples are generated from an environment simulator, which consists of V2V links and V2I links as well as their channel strengths. The vehicles are randomly dropped and the channel strengths for V2V and V2I links are generated based on the positions of the vehicles. With the selected spectrum and power of V2V links, the simulator can provide the next state, $s_{t+1}$, and the reward, $r_t$, to the agents. The training samples generated for optimizing the deep neural network consist of $s_t$, $s_{t+1}$, $a_t$, and $r_t$.

\begin{figure}[!t]
\centering
\includegraphics[width=\linewidth]{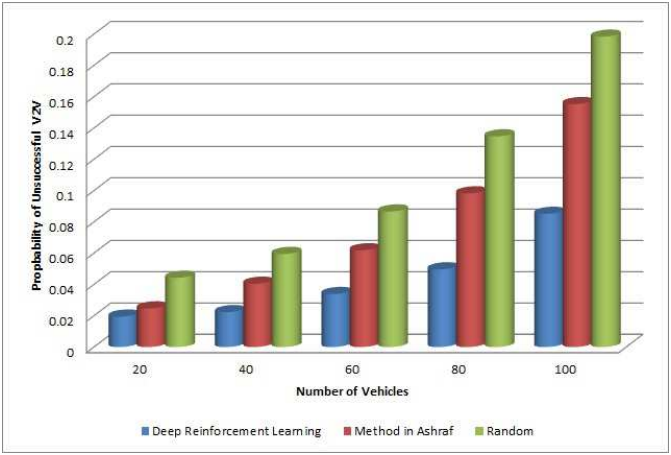}
\caption{Performance comparison of distributed resource management based on deep reinforcement learning.} \label{fig:V2V}
\end{figure}

The deep reinforcement learning based resource allocation scheme is compared with two methods in Fig.~\ref{fig:V2V}. The first is a random resource allocation method, where the agent randomly chooses a sub-band for transmission at each time. The other method is from \cite{ashraf2016dynamic}, where vehicles are first grouped into clusters according to the similarities of V2V links and then the sub-bands are allocated and adjusted iteratively in each cluster for better performance.
Fig.~\ref{fig:V2V} shows the probability that V2V links violate the latency constraint versus the number of vehicles. From the figure, the deep reinforcement learning method has a smaller probability for V2V links violating the latency constraint since it can dynamically adjust the power and sub-band for transmission so that the links that are likely to violate the latency constraint have more resources.

\section{Open Issues}

Even though remarkable progress has been made by machine learning in various areas, it is still insufficient to just naively apply the existing  learning algorithms in vehicular networks due to their distinctive characteristics.
In this section, we discuss several issues that need further attention.



\subsection{Method Complexity}
Unlike traditional machine learning techniques that require much effort on feature design, deep neural networks provide better performance by learning the features directly from raw data. Hence, information can be distilled more efficiently in deep neural networks than the traditional methods. It has been shown by experimental results that the deep hierarchical structure is necessary. Recently, in order to enhance the representation ability of the model, more advanced structures and technologies have since been devised, such as the LSTM as briefly discussed in Section~\ref{sec:dlIntro}. Moreover, with high-performance computing facilities, such as graphics processing unit (GPU), deep networks can be efficiently trained with massive amounts of data through advanced training techniques, such as batch norm \cite{ioffe2015batch} and residual networks \cite{he2016deep}.
However, computation resources aboard vehicles are rather limited and because of the stringent end-to-end latency constraints in vehicular networks, the use of powerful servers housed remotely for computation would also be confined. As a result, special treatments, such as model reduction or compression, should be carefully developed to alleviate the computation resource limitation without incurring much performance degradation.

\subsection{Distributed Learning and Multi-Agent Cooperation}
Different from most existing machine learning applications that assume easy availability of data, in vehicular networks, however, the data is generated and stored across different units in the network, e.g., vehicles, RSUs, remote clouds, etc.
As a consequence, distributed learning algorithms are desired such that they can act on partially observed data and meanwhile have the ability to exploit information obtained from other entities in the network.
Such scenarios can be technically modeled as a multi-agent system, where cooperation and coordination among participating agents play important roles in reaching system level optimal performance through sharing necessary information among each other.
Each individual vehicle agent thus gets more informed about the environment and jointly optimizes its performance with other agents in the network.
With machine learning, vehicle agents are able to learn what they need to share based on what they have perceived and what they need to do, with minimal network signaling overhead.

{In traditional multi-agent learning systems, the communications cost of message sharing among agents is not considered and the shared information is assumed to be error and delay free \mbox{\cite{foerster2016learning,Foerster2017stabilising}}. For high mobility vehicular networks, however, practical constraints imposed by the harsh environments should be considered. For example, the resources for communications, e.g., transmit power and bandwidth, are limited and the channel quality is time-varying. When the channel is in deep fading, received data suffer from severe errors \mbox{\cite{katabi}} and noticeable delay is also inevitable. As a result, developing efficient coordination and cooperation schemes for multiple vehicle agents while taking the wireless constraints into consideration needs to be further explored.}

\subsection{\textcolor{black}{{Security Issues}}}

{Machine learning has been shown to be helpful in confronting cyber-physical attacks, which threatens the safety of vehicular networks, as discussed in Section} \ref{sec:networkSecurity}.
Ironically, it also raises tremendous potential challenges and risks by itself since the machine learning based system can produce harmful or unexpected results \cite{AIsafety}. For instance, the convolutional neural networks can be easily fooled by maliciously designed noised images \cite{nguyen2015deep} while the agents in reinforcement learning may find undesirable ways to enhance the reward delivered by their interacting environment \cite{ring2011delusion}. As a result, even though machine learning has achieved remarkable improvement in many areas, significant efforts shall be made to improve the robustness and security of machine learning methods before they come to the safety-sensitive areas, such as vehicular networks, where minor errors may lead to disastrous consequences.

\subsection{\textcolor{black}{{Learning for Millimeter Wave Vehicular Networks}}}
\textcolor{black}{{The millimeter wave (mmWave) band is an attractive option to support high data rate communications for advanced safety and infotainment services in future vehicular networks with the availability of order of magnitude larger bandwidth \mbox{\cite{Choi2016millimeter,Va2016millimeter}}.
The small wavelength of mmWave bands makes it possible to pack massive antenna elements in a small form factor to direct sharp beams to compensate for the significantly higher power attenuation of mmWave propagation.
Over the past years, significant research efforts have been dedicated to addressing a wide range of problems in mmWave communications, including mmWave channel modeling, hybrid analog and digital precoding/combining, channel estimation, beam training, and codebook designs \mbox{\cite{Heath2016overview}}.}}

\textcolor{black}{{ A distinctive challenge of mmWave vehicular communications is the large overhead to train and point narrow beams to the right direction due to the constant moving of vehicles.
Besides, the mmWave transmission is susceptible to blockage and therefore fast and efficient beam tracking and switching schemes are critical in establishing and maintaining reliable mmWave links \mbox{\cite{Va2018inverse}}.
Machine learning tools can be effective in addressing such challenges, through exploiting historical beam training results \mbox{\cite{Alkhateeb2018}},
situational awareness \mbox{\cite{Wang2018mmWave}}, and other context information of the communications environment.
Mapping functions from the context information (features), such as environment geometry, network status, and user locations, to the beam training results can be learned using deep neural networks or other regression algorithms.
It remains to study the proper resolution levels for encoding/representing the context information to strike a balance between performance and computational complexity.
Moreover, it would be particularly interesting to see if more sophisticated machine learning models, such as RNNs and LSTMs to exploit temporal correlations, can achieve better performance in predicting mmWave beamforming directions in rapidly changing vehicular environments. }}

\section{Conclusion}
In this article, we have investigated the possibility of applying machine learning to address problems in high mobility vehicular networks.
Strong dynamics exhibited by such types of networks and the demanding QoS requirements challenge the state-of-the-art communications technologies.
Machine learning is believed to be a promising solution to this challenge due to its remarkable performance in various AI related areas.
We have briefly introduced the basics of machine learning and then provided some examples of using such tools to learn the dynamics and perform intelligent decision making in vehicular networks. We have further highlighted some open issues and pointed out areas that require more attention.


\bibliographystyle{IEEEtran}
\bibliography{ml_iot}

\end{document}